\documentclass[letter,12pt]{article}
\usepackage{amssymb}
\usepackage{amsfonts}
\usepackage{amsmath}
\usepackage{graphicx}
\usepackage{verbatim}

\title{Levy model of cancer}
\author{Roberto Herrero (1), Dario A. Leon (2,3) and Augusto Gonzalez (4,3)\\
(1) Faculty of Physics, University of Havana, Cuba\\
(2) University of Modena and Reggio Emilia, Italy\\
(3) Institute of Cybernetics, Mathematics and Physics,\\ 
 Havana, Cuba\\
(4) University of Electronic Sciences and Technology\\ 
 of China, Chengdu, People Republic of China}

\begin{document}

\maketitle

\begin{abstract}
A small portion of a tissue defines a microstate in gene expression space. Mutations, epigenetic
events or external factors cause microstate displacements which are modeled by combining small
independent gene expression variations and large Levy jumps, resulting from the collective
variations of a set of genes. The risk of cancer in a tissue is estimated as the microstate 
probability to transit from the normal to the tumor region in gene expression space. 
The formula coming from the contribution of large Levy jumps seems to provide a qualitatively correct
description of the lifetime risk of cancer, and reveals an interesting connection between
the risk and the way the tissue is protected against infections. 
\end{abstract}

\vspace{.5cm}

\begin{figure}[ht]
\begin{center}
\includegraphics[width=0.9\linewidth,angle=0]{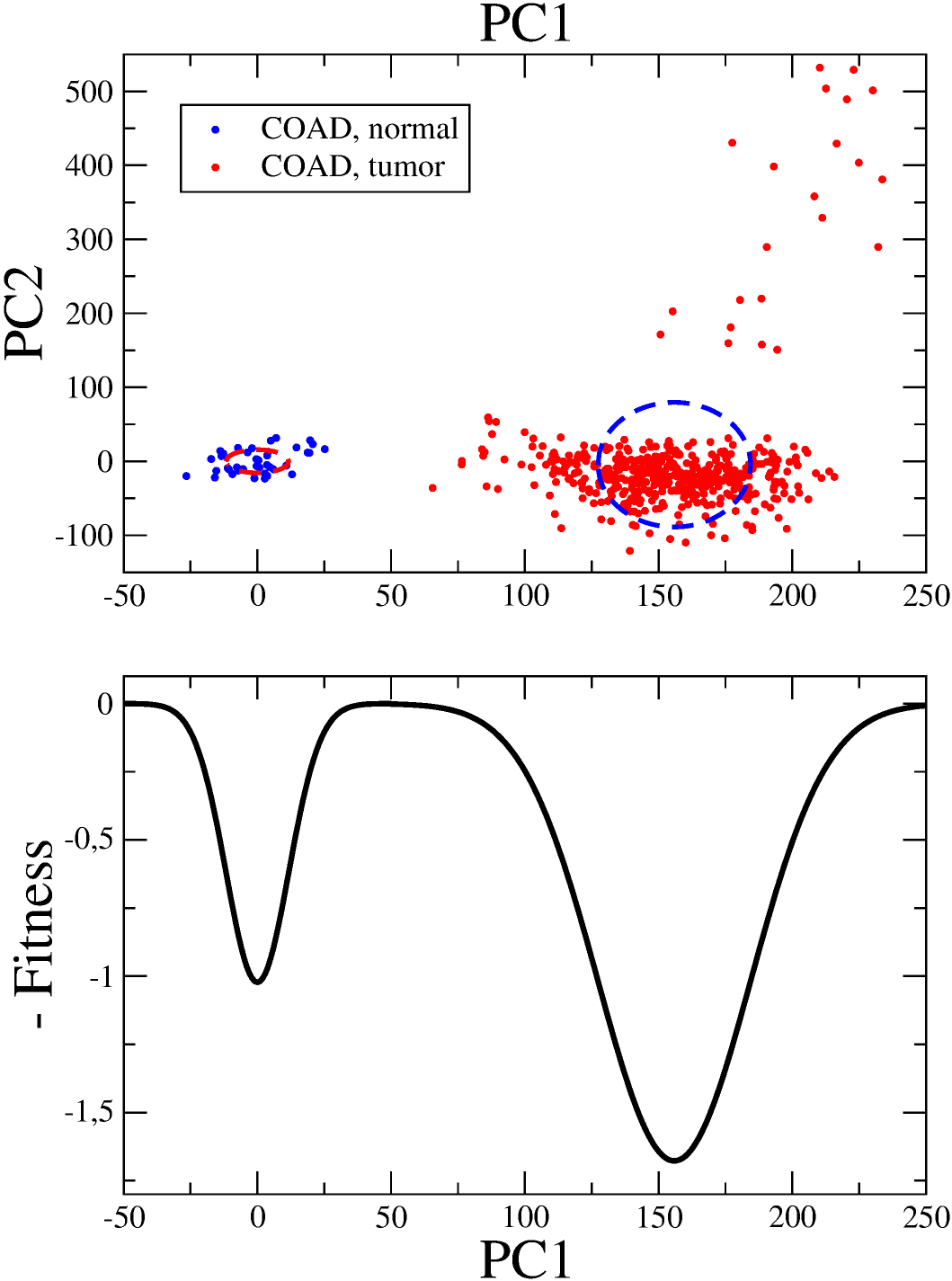}
\caption{Top: PC analysis of the GE data for adenocarcinoma of the colon. Normal (blue circles)
and tumor samples (red circles) are shown. Ellipses illustrating the centers and r.m.s. radi
of both clouds of points are drawn. Bottom: Schematics of the fitness landscape. The fitness is 
normalized to the homeostatic value. The tumor region exhibits the deepest well (highest fitness).}
\label{fig1}
\end{center}
\end{figure}

{\bf A 1D model for tumorigenesis.} 
We are closer than ever to modeling tumorigenesis. Perhaps, the best way to proceed is to use a 
gene expression (GE) description \cite{GEdescrip}. Indeed, in GE space the normal (homeostatic) and tumor states 
are seen as distant regions (attractors) \cite{landscape}. Recall, for example, 
the principal component analysis \cite{PCA1,PCA2,PCA3} of GE data for colon adenocarcinoma (COAD), obtained from The 
Cancer Genome Atlas portal (https://www.cancer.gov/tcga). Fig \ref{fig1} top panel shows the results \cite{landscape}.

Each point in this figure comes from a biopsy, Small samples are taken off from different patients
and processed in order to obtain expression values for 60483 genes. For each gene, we define a
reference value, $e_{ref}$, by geometric averaging over the normal samples. Then, new variables
are defined: $\hat e=\log _2(e/e_{ref})$. The origin of coordinates in this figure is precisely
the center of the cloud of normal samples, $\hat e=0$. A covariance matrix is defined and diagonalized.
The first eigenvectors are used to define new coordinate axes: PC1, PC2, etc. Details may be found 
in Ref. \cite{landscape}. We shall only stress that the first axis, PC1, which accounts for 51 \%
of the total data variance, is the cancer axis, which allows the discrimination between normal and 
tumor states.

Let ${\bf v_1}$ be the eigenvector of the covariance matrix along PC1, and ${\bf \hat e}$ the
expression vector corresponding to a given sample. Then, $x_1={\bf \hat e\cdot v_1}$ is the 
position along PC1 of the sample. Normal samples define a region around the origin with r.m.s.
radius $R_n=11.71$. On the other hand, the cloud of tumor samples is centered
at $\bar x_1=155.89$, and its r.m.s. radius is $R_t=28.53$ \cite{GErearr}.

Recall the interpretation of points in Fig. \ref{fig1} top panel. Each point comes from a
small sample, the GE data obtained from it contains the contribution of many cells
and the complex signaling system regulating their interactions. One may speak of a 
tissue microstate. On the other hand, points come from different patients, each carrying
a particular genetic load. The fact that the points are grouped in definite regions means
that these regions are indeed attractors in GE space.

We want to describe the genesis of a tumor, that is the time evolution of a portion or 
sample of a tissue that starts in the normal region and progress towards the tumor zone.
We have already defined a single coordinate describing this progression: $x_1$. In 
order to proceed further with the model, we shall clarify why and how this progression
takes place. 

The coordinate $x_1$ describing the tissue microstate starts at a point near the origin
and realizes random oscillations in the normal zone. The cause for such random displacements
is discussed in the next sections. The motion is confined to the normal
region for a long time  because this zone is a local maximum of fitness \cite{GErearr,entropy}.  We have
schematically represented in Fig. \ref{fig1} bottom panel the fitness distribution
along the PC1 axis. The $y$ axis of this figure is the fitness with a minus sign, thus
that the normal and tumor zones are local maxima. We have computed in Ref. \cite{entropy}
the number of available microstates in each zone, showing that this number is much greater
for tumors than for normal states. In other words, the volume of the basin of attraction
is much greater in the tumor than in the normal region. In addition, as a consequence of
breaking the restrictions imposed by homeostasis, the mitotic rate
of tumor stem cells is usually greater than that of normal somatic stem cells \cite{ratetumors}.
The conclusion is that the tumor minimum is the deepest in Fig. \ref{fig1}, the one with 
highest fitness. Our drawing for the fitness distribution is a sketch, but we are convinced that
at least a rough quantitative description can be reached.

The intermediate region, $R_n<x_1<\bar x_1-R_t$, holds a low-fitness barrier \cite{GErearr,entropy},
which prevents the spontaneous transitions from the normal to the tumor region. The relative 
scarcity of samples in this region evidences the existence of the barrier. 

A tissue microstate realizes random displacements within the normal region. Only when the barrier
is surpassed and the microstate leaves the normal basin of attraction it is driven towards
the tumor attractor. The transition is seen as discontinuous \cite{GErearr}.

A precise description of the transition requires the detailed knowledge of the fitness
landscape and the causes of the random fluctuations. However, in order to estimate the risk
of cancer in a tissue we may proceed in a simpler way and compute the probability for the 
$x_1$ variable describing the microstate to transit from the normal to the tumor region.
The minimal walk length is $\bar x_1-R_n-R_t$. This is the goal we are aimed at in the present paper.

The starting point in our model is a large set of samples or microstates located near
the origin of Fig. \ref{fig1}. They represent small portions of the healthy tissue.
We may think of colon crypts in the studied example. The mean number of crypts in a 
healthy individual is estimated as  $1.5\times 10^7$ \cite{crypts}. We shall follow 
the random oscillations in GE space of each of these crypts.

With regard to the time variable, it is natural to follow the renewal cycle of somatic
stem cells, guaranteeing crypt homeostasis. In the studied example, the renewal rate
is 73 per year \cite{ratecolon}. Thus, we shall measure time in terms of somatic
stem cell generations. $t=0$ may refer to conception or to the moment at which the 
first colon stem cell appears. On the other hand, $t_0=\log _2 N_{sc}$, where $N_{sc}$ 
is the number of stem cells in the tissue, is the moment at which the tissue is formed. 
In colon, $N_{sc}\approx 2\times 10^8$, and $t_0\approx 27$. This is our starting point. 
\vspace{.5cm}

{\bf Small random displacements in GE space.}
Small variations of GE levels spontaneously occurs and may have different origins. 
First, somatic mutations in the human genome are known to occur at a 
rate of 8 per cell generation \cite{somatic}. Second, there is also a rate of 
accumulation of epigenetic (mainly methylation and phosphorylation) events modifying
the normal expression levels \cite{epigenetic}. Both processes could be boosted by
inherited mutations \cite{inhermut1,inhermut2} or external carcinogens \cite{carcinogens}.

We may thus write for the $x_1$ coordinate, characterizing the microstate of a crypt
at time $t=n+1$, the following equation:

\begin{equation}
x_1^{(n+1)}=x_1^{(n)}+\delta x_1,
\label{eq1}
\end{equation}

\noindent
where 

\begin{equation}
\delta x_1={\bf v_1\cdot \delta\hat e}=\sum_{i}v_{1i}~\delta \hat e_i,
\label{eq2}
\end{equation}

\noindent
and $\delta \hat e_i$ corresponds to a random variation of the expression of gene i. Eq. (\ref{eq1})
describes a Markov chain of events \cite{Markov}. On the other hand, Eq. (\ref{eq2})
shows that fluctuations in the expression levels are filtered by the ${\bf v_1}$
vector.

\begin{figure}[ht]
\begin{center}
\includegraphics[width=\linewidth,angle=0]{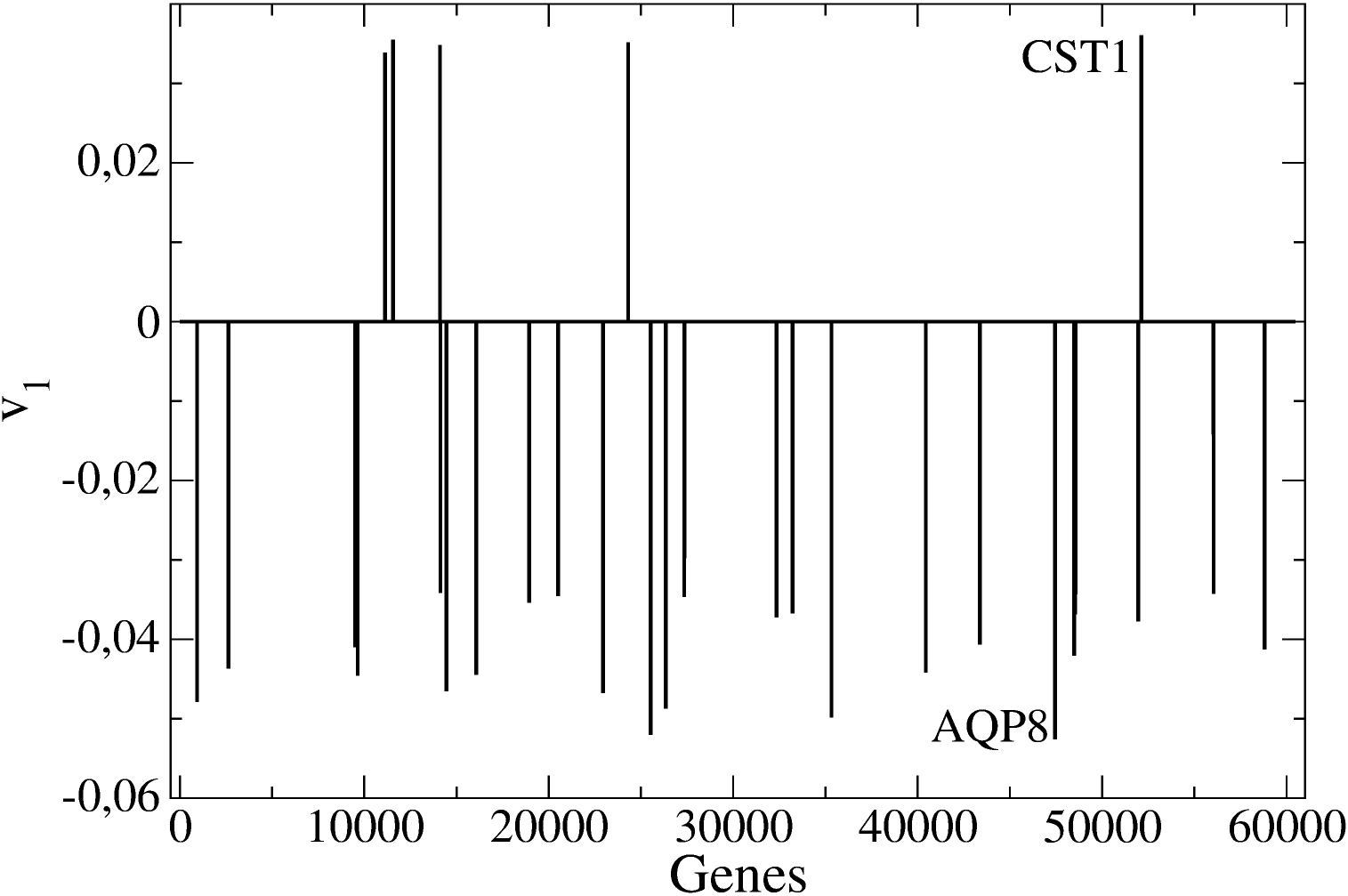}
\caption{The 30 genes with most significant contributions to the ${\bf v_1}$ vector 
in COAD. CST1 is stressed among the over-expressed and AQP8 among the silenced genes.}
\label{fig2}
\end{center}
\end{figure}

In Fig. \ref{fig2} we draw the 30 genes with the greatest contributions to ${\bf v_1}$ in COAD
\cite{landscape}. $v_{1i}>0$ and $v_{1i}<0$ correspond, respectively, to genes that should be 
over- and under-expressed (silenced) in order to the tumor to progress. We have distinguished 
the genes CST1 and AQP8. The former is a known marker of colon cancer \cite{CST1}, whereas the 
latter plays a significant role in colon homeostasis \cite{AQP8} and should be silenced in tumors.

The maximum value of $|v_{1i}|$ defines a scale, $D$, for the fluctuations of $x_1$. 
In COAD, it coincides with the modulus of the $v_{1i}$ related to the AQP8 gene. In order 
to get a simple estimate for the cancer risk, we may adopt the following model for the 
fluctuations: $\delta x_1=D~r$, where $r$ is a uniformly distributed random number in (-1,1).
This model may result from an independent variation hypothesis, i.e. random amplitudes and signs
in the individual gene variations $\delta \hat e_i$, so that most of they cancel.
In this way, Eq. (\ref{eq1}) for the small displacements in GE space describes a 1D Brownian
or Poisson process \cite{Brownian}.

We may use the well known fact that in a Brownian process, the final amplitudes at a given 
time are normally distributed, i.e. the probability density is given by:

\begin{equation}
p(x)=\sqrt{a/\pi}~ e^{-a (x-x_0)^2},
\label{eq3}
\end{equation}

\noindent
where $a=2/(D^2 t)$. We shall evaluate the probability for a trajectory starting in the normal zone
to reach the tumor zone. Above, we pointed out that the minimal walk length is $R=\bar x_1-R_n-R_t$. 
Thus, an estimate for the risk may be obtained from:

\begin{equation}
\int_{R}^{\infty} p(x) {\rm d}x=Err(\sqrt{a R^2}),
\label{eq4}
\end{equation}

\noindent
where $Err$ is the complementary error function. The argument of this function is $z=\sqrt{a R^2}=\sqrt{2/t}~R/D$,
in principle a large number. Then, we may use the asymptotic behavior $Err(z)\approx 
\exp(-z^2)/(\sqrt{\pi}z)$ for large $z$. The risk of cancer in COAD is obtained by multiplying
the escape probability for a single crypt by the number of crypts, or by the number of stem cells,
which is proportional to it:

\begin{equation}
risk \sim N_{sc}~ \frac{D\sqrt{t}}{R}~e^{-2(R/(D\sqrt{t}))^2},
\label{eq5}
\end{equation}

\noindent
or

\begin{equation}
ln (risk/N_{sc}) = const + ln(D\sqrt{t}/R)-2(D\sqrt{t}/R)^{-2}.
\label{eq6}
\end{equation}

This expression is general enough to be applied to other tissues, besides colon.
The constant in Eq. (\ref{eq6}) may account for other effects as, for example, the
role of the immune system. Microregions escaping the normal region and forming
a proto-tumor could be the subject of an attack by the immune system in the very
early stages \cite{immune}. By definition, the constant is less than zero because
the overall constant in Eq. (\ref{eq5}) is less than one.

In Table \ref{tab1} we compile a set of parameters for a group of tumors.
The geometry of the normal and tumor regions, i.e. the parameters $\bar x_1$,
$R_n$ and $R_t$ come from Ref. \cite{GErearr}. The $D$ value is estimated
as the maximum of $|v_{1i}|$ \cite{landscape}. On the other hand, the
number of tissue stem cells, $N_{sc}$, the stem cell turnover rate. $m_{sc}$,
and the lifetime risk of cancer (when available) are borrowed from Refs.
\cite{Tomasetti1,Tomasetti2}. The reported values of risk represent averages
over 380 cancer registries from different cities and countries in the world
\cite{Tomasetti2}.

We may test Eq. (\ref{eq6}) for the risk of cancer in a tissue resulting from 
small random variations of GE levels by using the data included in Table \ref{tab1}.
A plot of the l.h.s. vs the r.h.s. of Eq. (\ref{eq6}) should lead to a straight
line with a slope near one and a constant less than zero. Notice that the life 
expectancy in Ref. \cite{Tomasetti1} is assumed to be 80 years. Thus, $t$ is
obtained by multiplying the stem cell rate, $m_{sc}$, by 80 years.

\begin{figure}[ht]
\begin{center}
\includegraphics[width=\linewidth,angle=0]{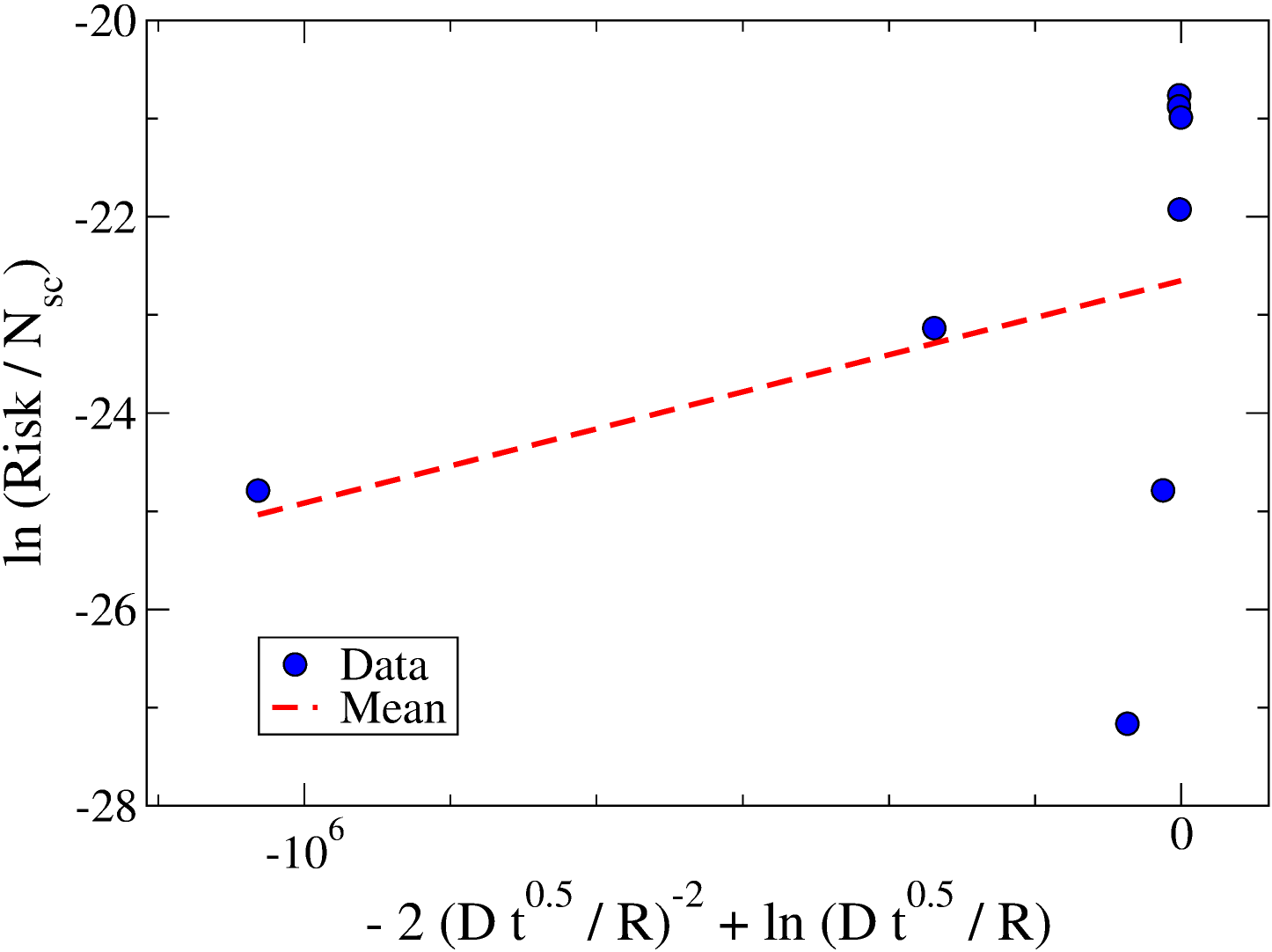}
\caption{A test of how Eq. (\ref{eq6}) describes the risk of cancer in 8 tissues.
Data from Table \ref{tab1} is used to this end. A nearly flat dependence on the 
abcisa is obtained, thus small amplitude fluctuations in gene expression space
may not account for the risk of cancer in these tissues..}
\label{fig3}
\end{center}
\end{figure}

The results of that test are shown in Fig. \ref{fig3}. We get a nearly flat
curve (slope = $2\times 10^{-6}$), indicating that the proposed dependence of the risk 
on the parameters is not correct. Thus, the 
observed risk of cancer can not be explained by small amplitude random variations
in GE values. In the next section, we shall consider large jumps.

Let us stress that we use an expression like $t=m_{sc}\times age$ in a very broad age
interval. It is well known that $m_{sc}$ experiences a significant decrease
as a result of aging \cite{aging1,aging2}. However, also as a consequence of aging there is an 
accumulation of epigenetic events and DNA damages leading to a reduction of 
fitness and a displacement towards the low-fitness zone. Thus, aging acts
in the same direction as the low amplitude fluctuations of GE values.

\vspace{.5cm}

{\bf Large (Levy) jumps in GE space.}
Besides small random displacements, related to quasi independent variations in the GE
values, there is also the possibility of large jumps in GE space. The origin of such 
large motions could be diverse.

First, there are large scale mutations, involving DNA rearrangements and simultaneously
modifying the expression of many genes. An example, known to play an important role
in cancer, is that of aneuploidies \cite{aneup}.

Second, large jumps in GE space could be related to coordinated variations in a group
of genes. Indeed, GE values are known to be regulated by GE networks \cite{networks}.
The global states of these networks define attractors. Variations in genes playing a
decisive role in the network, or accumulation of variations in many genes, may cause a
transition from one of these global states to another one.

Third, there is also the possibility of a programmed chain of mutations and GE
variations leading to cancer, triggered by let unknown causes, which is the basic
hypothesis in the atavistic theory of cancer \cite{atavistic}.

For the large GE variations, we shall specify their rate of occurrence, $\mu$, and
the probability distribution for their amplitudes, $\pi (\Delta x_1)$.

It is very plausible to assume that $\pi$ is of Pareto \cite{Pareto} or Levy \cite{Levy}
kind, with a power-like tail. Indeed, the Pareto character of GE distribution functions
was demonstrated in Ref. \cite{ParetoGE} (see also \cite{GErearr}). The Levy character
of the length distribution functions in mutations was shown in \cite{Levymut}.

Thus, our assumption is that displacements in GE space are a kind of Levy flights. 
Small variations allow the exploration of the fitness landscape at lower scales,
whereas sporadic large jumps allow to find global maxima. Besides mutations \cite{Levymut},
Levy flights are known to take place in many other biological processes, for example
foraging \cite{foraging}.

For lage $|\Delta x_1|$, the tail of $\pi$ is described by a Pareto exponent $\nu$:

\begin{equation}
\pi (\Delta x_1)\sim 1/|\Delta x_1|^\nu.
\label{eq7}
\end{equation}

The probability of a large jump reaching the tumor region is thus
proportional to 

\begin{equation}
D ~(\mu t)~ \int_{R}^{\infty} {\rm d}x/x^\nu,
\label{eq8}
\end{equation}

\noindent
and the risk of cancer in a tissue:

\begin{equation}
risk\sim N_{sc}D~ \mu~ (t_0+m_{sc}\times age)/R^{\nu-1},
\label{eq9}
\end{equation} 

\noindent
where we assume $\nu>1$. Below, we use $\nu=2$ in order to get an estimate of the risk.

Let us examine Eq. (\ref{eq9}) in more details. First, Eq. (\ref{eq8}) assumes that $R$ is in 
the tail of the distribution function. This is justified if we compare $R=115.65$ with the
scale $D=0.0526$. Second, no more than one hit or large jump is assumed to occur in the 
evolution of each microstate. In other words, the probability  $\mu t$ is less than one
and large jumps are thought to be rare. Third, we should consider the possibility of 
large GE variations due to mutations in the development period, that is why we included 
$t_0=\log_2(N_{sc})$ in the formula. This
is particularly important in tissues with slow renewal rates but large number of stem
cells. For example, in lung $t_0\approx 30$, but $m_{sc}\times 80$ years is only 5.6. Fourth,
the rate of large jumps, $\mu$, is unknown. However, if we assume roughly the same value for
all tissues, then it can be absorbed in the overall constant entering Eq. (\ref{eq9}).
The Pareto exponent is also unknown. Notice that in the COAD GE distribution functions 
the exponents take values between 1.6 and 2.0 \cite{GErearr}. The value we use for estimates, 
$\nu= 2$, is motivated by this result.

Finally, we get the following expression for the risk, which may be tested against the
data in Table \ref{tab1}:

\begin{equation}
ln (risk/N_{sc})=const + ln \{D(t_0+m_{sc}\times age)/R\}.
\label{eq10}
\end{equation}

\begin{figure}[ht]
\begin{center}
\includegraphics[width=\linewidth,angle=0]{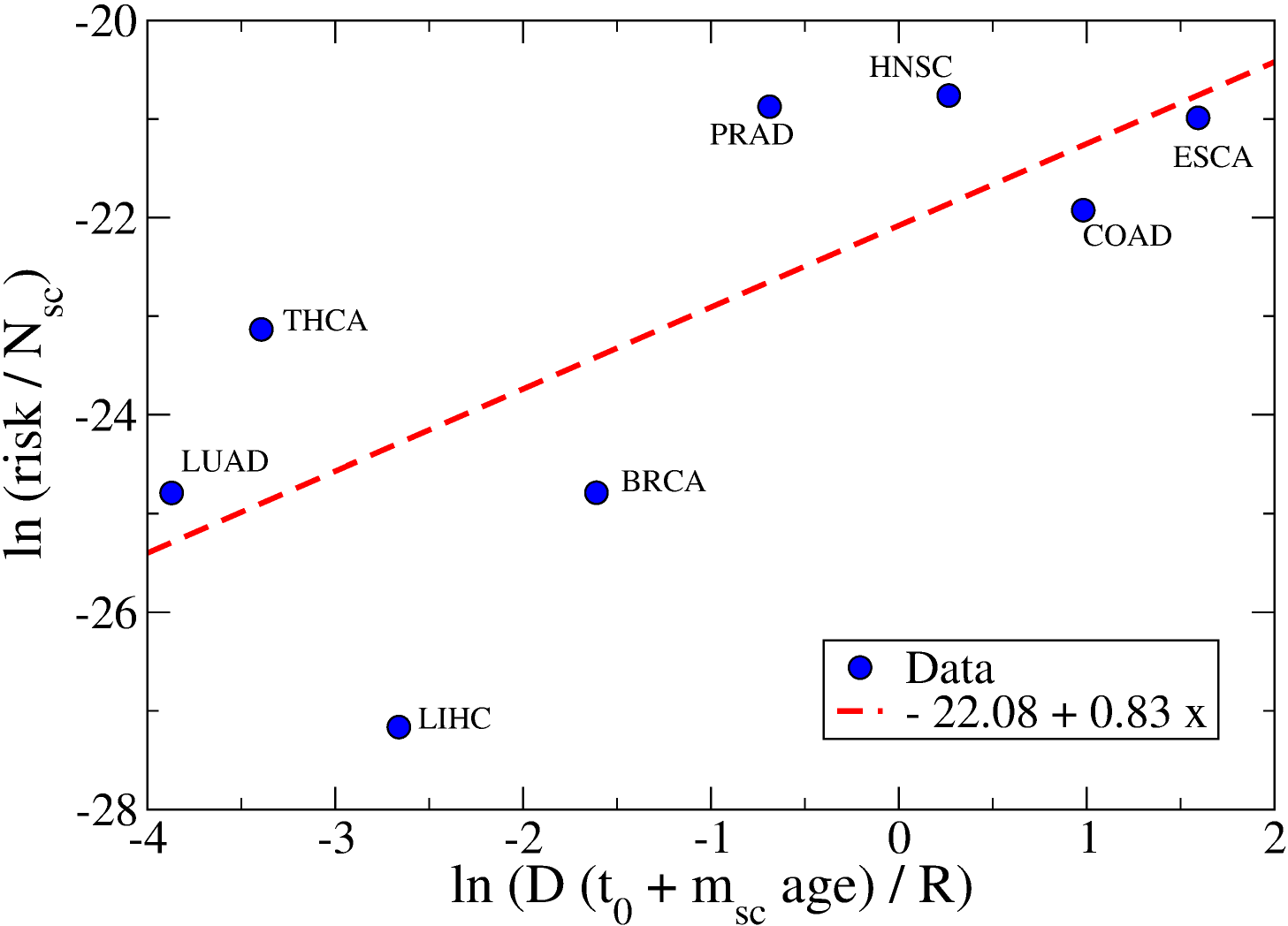}
\caption{Testing the ability of Eq. (\ref{eq10}) to describe the 
cancer risk in 8 tissues. The slope of the linear fit is near one, as expected.
72 \% of the data dispersion is explained by the linear dependence (p-value = 0.04).}
\label{fig4}
\end{center}
\end{figure}

\noindent
The constant should be negative according to our hypothesis of $\mu$ small. The results of the test are shown in Fig. \ref{fig4}.

The observed behavior is consistent with a linear dependence with slope near one.
The Pearson correlation coefficient is 0.85, with a p-value equal to 0.04. 
The main reason for the remaining dispersion of points could be the assumption that
the rate of large jumps, $\mu$, is roughly the same for all of the tumors. A variable 
$\mu$ would account for the dispersion.

In conclusion, we get the following simple expression for the risk of cancer per
stem cell in a tissue, coming from large jumps in GE space:

\begin{equation}
\frac{risk}{N_{sc}}=\mu' \frac{D}{R} (t_0+m_{sc}~age),
\label{eq11}
\end{equation}

\noindent
where we included an effective rate, $\mu'$. Genetic, viral or external carcinogenic factors
may increase $\mu'$, whereas the action of the immune system in the tissue may modify
$\mu'$ in any direction. In the next section, we qualitatively analyze a larger set
of tumors by using Eq. (\ref{eq11}). 
\vspace{.5cm}

{\bf Qualitative analysis of the data on cancer risk in different tissues.}
We use Eq. (\ref{eq11}) in order to re-examine the data presented in paper \cite{Tomasetti1}. 
The qualitative idea is to set a reference value for the coefficient in front
of the r.h.s. of Eq. (\ref{eq11}), $a_{ref}=2\times10^{-14}$, and include an
extra risk score (ERS), in such a way that:

\begin{equation}
\frac{risk}{N_{sc}}=ERS~ a_{ref}~ (t_0+m_{sc} age).
\label{eq12}
\end{equation}

\begin{figure}[ht]
\begin{center}
\includegraphics[width=\linewidth,angle=0]{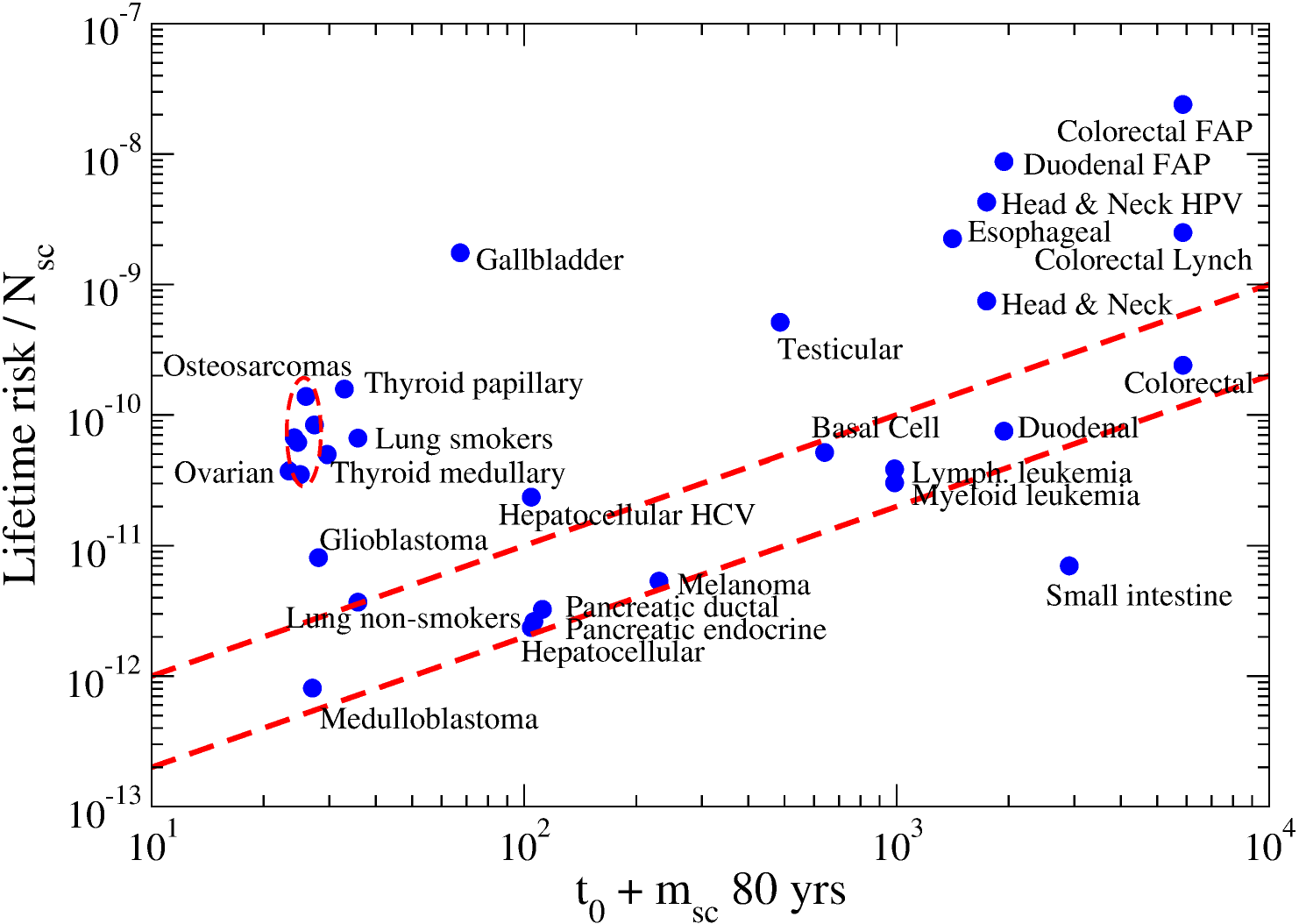}
\caption{(Color online) Lifetime risk of cancer per stem cell in a tissue vs the number of 
stem cell generations. The analysis is based on Eq. (\ref{eq12}). See the explanation in the main text.}
\label{fig5}
\end{center}
\end{figure}

We should try to understand the observed values of ERS in terms of the tissue
characteristics. The results are shown in Fig. \ref{fig5}. In
order to facilitate the analysis, the studied tumors are separated in groups.

Group I includes 11 tumors (9 tissues), located in a band delimited by red dashed lines 
in Fig. \ref{fig5}, and coefficients $1<ERS<5$. In the lack of a better name, it is called 
the normal group. In this set, random fluctuations in GE space seem to play the main
role in the genesis of cancer, as originally claimed in Ref. \cite{Tomasetti1}. 
Notice that this group is conformed by 
very different tissues -- from the medulloblastoma to the colorectal adenocarcinoma.

Group II, with five points in the figure, include cases in which genetic or viral causes 
enhance the rate $\mu'$. The ERS index exhibits very high values in this set.

The abnormal values of ERS for the 7 tissues (12 points) contained in Group III could have
an immunological origin. Indeed, our body uses physical barriers, humors and immune
cells in order to protect the tissues against infections caused by pathogens, which are the most common
attacks. The combined effects of these factors guarantees immunity. In tissues where one
factor is predominant, the others could be somehow depressed. On the other hand,
the protection against tumors, which come form inside, that is originate in tissue cells,
is mainly the responsibility of immune cells.

Barriers are known to play a basic role in the protection of germinal cells \cite{Blood-Testis} 
and the brain \cite{Blood-Brain} against infections.
The cellular component of immunity in these tissues is, 
in some way, depressed with the purpose of avoiding inflammation events. 
The relatively high values of ERS could be explained in this way. 

By contrast, the inclusion of the Medulloblastoma in the normal group is probably related to
regional differences in blood-brain barrier permeability \cite{Cerebellum}.

With regard to bones, it is known that immunity relies strongly on defensins \cite{Bones}, 
possibly with a depressed role of immune cells. On the other hand, 
the thyroid is known to have a close cross-talk with the immune system \cite{Thyroid}. It's
dysregulation is the cause of immune disorders. One may speculate that a low cellular
response is needed in order to prevent dysregulation of the thyroid.

The extreme case in this group is gallbladder non-papillary adenocarcinoma, with an index $ERS=1300$, 
the understanding of which is a real challenge. However, one can speculate that the cellular response
is also depressed in the gallbladder, because of the strong microbicide character of the bile \cite{Bile}.

On the other hand, the relatively low value of ERS for 
the small intestine adenocarcinoma (eight times lower than the reference) can not have other explanation than 
overprotection by the cellular component of the immune system. Indeed, the small intestine is a possible
entrance door for the microbiota living in the colon, and as such it requires special protection.
The mean value of microbes/gm experiences a jump from $10^4$ to $10^{11}$ as we cross from the ileum 
to the cecum \cite{microbes}. Barriers can not be reinforced because of the reduced dimensions.
Thus, perhaps the Paneth cells \cite{Paneth}, Peyer's patches \cite{Peyer}, 
and other structures concentrated in the distal ileum are the  
responsible for this additional protection. 

Finally, there is a group of 3 tissues exhibiting abnormally high values of the ERS index,
presumably related to external factors. One example is lung adenocarcinoma, for which the concurrence of
radioactive Radon and smoking produces a 90-fold increase of the slope. 
\vspace{.5cm}

{\bf Concluding remarks.}
In the present paper, the time evolution of microstates representing small portions of
a tissue are described as Levy flights in gene expression space. The small amplitude Brownian
component is characterized by a radius $D \sqrt{t}$, much less than the distance between the
normal and tumor regions, $R=\bar x_1-R_n-R_t$. Only sporadic large jumps, of Levy nature, allow the microstate to
reach the cancer basin of attraction.

Although it is understood that aging induces a motion in the direction of the low-fitness
region, it was not explicitly included in our model. Work along this direction is necessary.

The resulting formula for the risk of cancer in a tissue was quantitatively tested against 
observed data, and applied to the qualitative analysis of the risk of cancer in a set of tissues. 
The most important conclusion, in our opinion, is a possible connection between the risk and
the way the tissue is protected against infections. The blood-brain barrier in the cerebrum, for
example, preventing the entrance of pathogens, is also the reason for the relatively low 
rate of elimination of prototumors, and thus large risk per stem cell in this organ.
The low risk per stem cell in the small intestine, on the other hand, is understood as a 
reinforcement of the cellular component of immunity.
\vspace{.5cm}

{\bf Acknowledgments.}
A.G. acknowledges the Cuban Program for Basic Sciences, the Office of External Activities of 
the Abdus Salam Centre for Theoretical Physics, and the University of Electronic Science and 
Technology of China for support. The research is carried on under a  project of the Platform 
for Bioinformatics of BioCubaFarma, Cuba. 

\bibliographystyle{unsrt}
\bibliography{biblio}

\begin{comment}

\end{comment}

\newpage

\begin{table}[t]
  \begin{center}
    \begin{tabular}{|l|l|l|l|l|l|l|l|l|} 
      \hline
      \textbf{Tissues} & $\bar x_1$ & $R_n$ & $R_t$ & $R$ &
 $D$ & $N_{sc}$ & $m_{sc}$ (1/yr) & risk \\
      \hline
      \hline
      BRCA & 137.37 & 20.97 & 31.66 & 84.74 & 0.0450 & $6.5\times 10^9 $ & 4.3 & 0.1116 \\
      \hline
      COAD & 155.89 & 11.71 & 28.53 & 115.65 & 0.0526 & $2\times 10^8$ & 73 & 0.0601 \\
      \hline
      HNSC & 123.50 & 27.74 & 23.54 & 72.22 & 0.0549 & $1.85\times 10^7$ & 21.15 & 0.0178 \\
      \hline
      KIRC & 171.80 & 28.70 & 36.00 & 107.1 & 0.0679 & ... & ... & ... \\
      \hline
      KIRP & 163.42 & 19.90 & 27.78 & 115.74 & 0.0768 & ... & ... & ... \\
      \hline
      LIHC & 134.67 & 20.48 & 45.23 & 68.96 & 0.0461 & $3.01\times 10^9$ & 0.9125 & 0.0048 \\
      \hline
      LUAD & 145.33 & 13.51 & 32.06 & 99.76 & 0.0581 & $1.22\times 10^9$ & 0.07 & 0.0209 \\
      \hline
      LUSC & 194.49 & 11.62 & 36.65 & 146.22 & 0.0522 & ... & ... & ... \\
      \hline
      PRAD & 91.33 & 31.31 & 32.17 & 27.85 & 0.0523 & $2.1\times 10^8$ & 3 & 0.1804 \\
      \hline
      STAD & 136.97 & 27.14 & 43.24 & 66.59 & 0.0455 & ... & ... & ... \\
      \hline
      THCA & 112.54 & 20.02 & 39.85 & 52.67 & 0.0532 & $8.25\times 10^7$ & 0.087 & 0.0074 \\
      \hline
      UCEC & 171.38 & 38.24 & 22.14 & 111.00 & 0.0439 & ... & ... & ... \\
      \hline
      BLCA & 140.61 & 57.53 & 34.68 & 48.40 & 0.0512 & ... & ... & ... \\
      \hline
      ESCA & 138.70 & 64.28 & 35.79 & 38.63 & 0.0710 & $6.65\times 10^6$ & 33.18 & 0.0051 \\
      \hline
      READ & 168.05 & 22.90 & 28.81 & 116.34 & 0.0521 & ... & ... & ... \\
      \hline
    \end{tabular}
 \caption{A set of parameters compiled for a group of tumors.
The geometry of the normal and tumor regions, i.e. the parameters $\bar x_1$,
$R_n$ and $R_t$ come from Ref. \cite{GErearr}. The minimal distance between both
regions is $R=\bar x_1-R_n-R_t$. The $D$ value is estimated
as the maximum of $|v_{1i}|$ \cite{landscape}. On the other hand, the
number of tissue stem cells, $N_{sc}$, the stem cell turnover rate. $m_{sc}$,
and the lifetime risk of cancer (when available) are borrowed from Refs.
\cite{Tomasetti1,Tomasetti2}.}
    \label{tab1}
     \end{center}
\end{table}

\newpage

\begin{table}[ht]
\begin{center}
\begin{tabular}{|c|c|}
\hline
 Cancer type & ERS \\
 \hline
 {\bf Group I. Normal} &\\
 {Hepatocellular C} & 1.13 \\
 {Melanoma} & 1.16 \\
 {Pancreatic endocrine C} & 1.23 \\
 {Pancreatic ductal AC} & 1.45 \\
 {Medulloblastoma} & 1.49 \\
 {Myeloid leukemia} & 1.54 \\
 {Duodenal AC} & 1.93 \\
 {Lymphocytic leukemia} & 1.95 \\
 {Colorectal AC} & 2.04 \\
 {Basal Cell C} & 4.02 \\
 {Lung AC (non-smokers)} & 5.15 \\
\hline
 {\bf Group II. Viral and Genetic} &\\
 {Hepatocellular C with HCV} & 11.29 \\
 {Colorectal AC with Lynch} & 21.30 \\
 {Head and Neck SCC with HPV} & 122.96 \\
 {Colorectal AC with FAP} & 204.51 \\
 {Duodenal AC with FAP} & 225.29 \\
\hline
 {\bf Group III. Immune} &\\
 {Small intestinal AC} & 0.12 \\
 {Glioblastoma} & 14.48 \\
 {Testicular germinal cell} & 52.78 \\
 {Ovarian germinal cell} & 79.86 \\
 {Thyroid medullary C} & 84.22 \\
 {Osteosarcomas} & 153.04 \\
 {Thyroid papillary and follicular C} & 239.78 \\
 {Gallbladder non papillary AC} & 1299.58 \\ 
\hline
 {\bf Group IV. Abnormal} &\\ 
 {Head and Neck SCC} & 21.38 \\
 {Esophageal SCC} & 79.44 \\
 {Lung AC (smokers)} & 92.77 \\
\hline
\end{tabular}
\caption{The Extra Risk Score (ERS) index of Eq. (\ref{eq12})
for cancer in different tissues.}
\label{tab2}
\end{center}
\end{table}

\newpage

\newpage

\newpage

\newpage

\newpage

\end{document}